\documentstyle[aps,twocolumn,epsf,psfig]{revtex}

\begin{document}
\newtheorem{theorem}{Theorem}
\newtheorem{acknowledgement}[theorem]{Acknowledgement}
\newtheorem{algorithm}[theorem]{Algorithm}
\newtheorem{axiom}[theorem]{Axiom}
\newtheorem{claim}[theorem]{Claim}
\newtheorem{conclusion}[theorem]{Conclusion}
\newtheorem{condition}[theorem]{Condition}
\newtheorem{conjecture}[theorem]{Conjecture}
\newtheorem{corollary}[theorem]{Corollary}
\newtheorem{criterion}[theorem]{Criterion}
\newtheorem{definition}[theorem]{Definition}
\newtheorem{example}[theorem]{Example}
\newtheorem{exercise}[theorem]{Exercise}
\newtheorem{lemma}[theorem]{Lemma}
\newtheorem{notation}[theorem]{Notation}
\newtheorem{problem}[theorem]{Problem}
\newtheorem{proposition}[theorem]{Proposition}
\newtheorem{remark}[theorem]{Remark}
\newtheorem{solution}[theorem]{Solution}
\newtheorem{summary}[theorem]{Summary}    
\def\r{{\bf{r}}}
\def\i{{\bf{i}}}
\def\j{{\bf{j}}}
\def\m{{\bf{m}}}
\def\k{{\bf{k}}}
\def\kt{{\tilde{\k}}}
\def\mt{{\hat{t}}}
\def\mG{{\hat{G}}}
\def\mg{{\hat{g}}}
\def\mGa{{\hat{\Gamma}}}
\def\mS{{\hat{\Sigma}}}
\def\mT{{\hat{T}}}
\def\K{{\bf{K}}}
\def\P{{\bf{P}}}
\def\q{{\bf{q}}}
\def\Q{{\bf{Q}}}
\def\p{{\bf{p}}}
\def\x{{\bf{x}}}
\def\X{{\bf{X}}}
\def\Y{{\bf{Y}}}
\def\F{{\bf{F}}}
\def\G{{\bf{G}}}
\def\bG{{\bar{G}}}
\def\mbG{{\hat{\bar{G}}}}
\def\M{{\bf{M}}}
\def\V{\cal V}
\def\tchi{\tilde{\chi}}
\def\tx{\tilde{\bf{x}}}
\def\tk{\tilde{\bf{k}}}
\def\tK{\tilde{\bf{K}}}
\def\tq{\tilde{\bf{q}}}
\def\tQ{\tilde{\bf{Q}}}
\def\si{\sigma}
\def\ep{\epsilon}
\def\hep{{\hat{\epsilon}}}
\def\al{\alpha}
\def\be{\beta}
\def\ep{\epsilon}
\def\bep{\bar{\epsilon}_\K}
\def\up{\uparrow}
\def\de{\delta}
\def\De{\Delta}
\def\up{\uparrow}
\def\dwn{\downarrow}
\def\ksi{\xi}
\def\etha{\eta}
\def\product{\prod}
\def\goto{\rightarrow}
\def\switch{\leftrightarrow}
                           
\title{Analysis of the dynamical cluster approximation for the  Hubbard
model}
\author{
K. Aryanpour$^{1}$
\and    
M. H. Hettler$^{2}$ and M. Jarrell$^{1}$ 
}
\address{$^{1}$University of Cincinnati, Cincinnati OH 45221, USA}
\address{$^{2}$Forschungszentrum Karlsruhe, Institut f\"ur Nanotechnologie,
 Karlsruhe, Germany}

\maketitle

\begin{abstract}
We examine a central approximation of the recently introduced 
Dynamical Cluster Approximation (DCA) by example of the Hubbard 
model. By both analytical and numerical means we study non--compact 
and compact contributions to the thermodynamic potential. We show
that approximating non--compact diagrams by their cluster analogs 
results in a larger systematic error as compared to the compact 
diagrams.  Consequently, only the compact contributions should be 
taken from the cluster,  whereas non-compact graphs should be 
inferred from the appropriate Dyson equation. The distinction 
between non--compact and compact diagrams persists even in the 
limit of infinite dimensions.  Non-local corrections beyond the DCA 
exist for the non--compact diagrams, whereas they vanish for compact 
diagrams.
\end{abstract}

\paragraph*{Introduction}
Strongly correlated electron systems are often characterized by short 
range dynamical fluctuations. Consequently, local approximations like 
the dynamical mean field approximation 
(DMFA)\cite{metzner-vollhardt,muller-hartmann,review1,review2} successfully 
describe many of the qualitative properties. However, in low--dimensional 
systems, spatial correlations become increasingly important and are thought 
to be responsible for e.g. non-Fermi-liquid behavior and d-wave pairing 
in the cuprate superconductors.

The Dynamical Cluster Approximation (DCA) was introduced as a technique 
to include such non-local corrections to DMFA\cite{DCA_hettler,DCA_hettler2}.  
This is accomplished by mapping the lattice problem onto that of a 
self-consistently embedded cluster, with periodic boundary conditions.  The 
DCA may also be viewed diagrammatically as an approximation which 
systematically restores momentum conservation at the internal vertices of 
many body Feynman diagrams, which is relinquished in the local DMFA.  Here, 
we investigate one of the central approximations of the DCA: that compact 
(skeletal) contributions to the thermodynamic potential are well 
approximated by their cluster counterparts; whereas non-compact 
(non-skeletal) contributions should not be approximated by their cluster 
counterparts.  Rather, they should be constructed using the 
appropriate Dyson equation.
 
\begin{figure}
\epsfxsize=3.5in
\hspace*{0.01cm}\epsffile{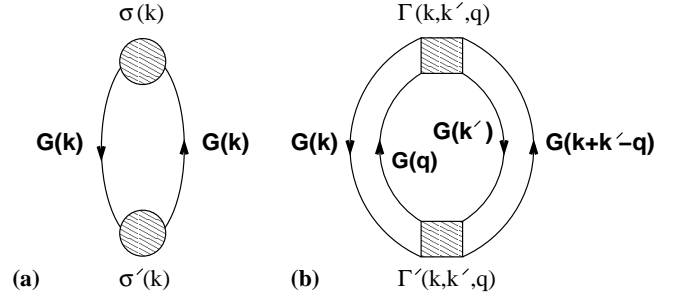}
\caption[a]{\em{(a) typical non-compact (non skeletal) and 
(b) typical compact (skeletal) diagrams.}}
\label{comp_noncomp}
\end{figure}

\paragraph*{General considerations}
Following Baym\cite{Baym}, a microscopic theory may be defined 
by its approximation to the generating functional $\Phi[G]$ defining the
thermodynamic potential (difference) of the system via
\begin{equation}
\label{eq:thermPot}
\Delta\Omega=\Omega-\Omega_0 = -2 {\rm Tr} [\Sigma G - {\rm ln}
(G/G_{0})] + \Phi[G]\,,
\end{equation}
where G is the full and $G_0$ the bare one particle Green function, and 
$\Sigma$ the self energy.  $\Phi[G]$ is a sum of all compact (skeletal) 
closed connected Feynman diagrams.  The other contributions to the 
thermodynamic potential incorporate non-compact diagrams. Typical 
compact and non-compact diagrams are illustrated in Fig.~\ref{comp_noncomp}. 
In the non-compact diagram, two self energy pieces $\sigma$ and $\sigma'$ 
are connected with two Green functions. In the compact diagram, two 
vertex parts, $\Gamma$ and $\Gamma'$, are connected with four Green 
functions.

As shown by M\"uller--Hartmann\cite{muller-hartmann}, the DMFA may be 
defined by relinquishing the momentum conservation at each internal 
vertex in  $\Phi[G]$\cite{note}.  This conservation is described by 
the  Laue function 
\begin{equation}
\label{eq:Laue}
\De=\sum_\x e^{i\x\cdot(\k_1+\k_2-\k_3-\k_4)}=N\de_{\k_1+\k_2,\k_3+\k_4}\,.
\end{equation}
In the Dynamical Mean Field Approximation (DMFA), momentum conservation 
is completely abandoned and the Laue function $\De_{DMFA}\equiv 1$. We 
may then sum freely over all the internal momenta labels.

\begin{figure}
\epsfxsize=3.3in
\epsffile{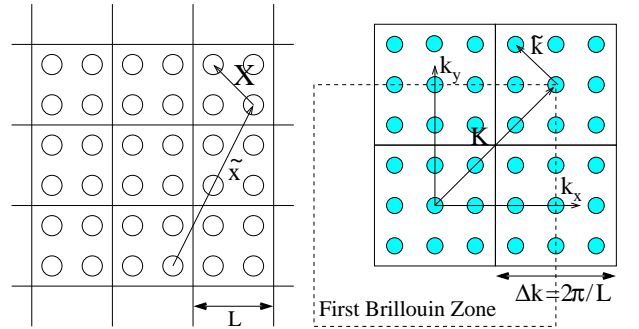}
\caption[a]{\em{The real lattice cluster (left) and the first
Brillouin zone (left) divided into subcells.}}

\label{divide_x_k}
\end{figure}
        The DCA is constructed to systematically restore the momentum
conservation at each internal vertex, by mapping the lattice onto
a self-consistently embedded cluster problem.  We have provided a
microscopic definition of the DCA through its Laue function.  However,
to clarify the relation between this microscopic definition, and the
cluster problem, we must first decompose the lattice into clusters,
and define the corresponding problem in reciprocal space.  Here, the 
real lattice of $N$ sites is tiled by $N/N_{c}$ clusters each composed 
of $N_{c}=L^D$ sites where D is dimensionality and $L$ the size of 
clusters (c.f.\ Fig.~\ref{divide_x_k} for $L=D=2$).  We label the origin 
of the clusters by $\tx$ and the $N_{c}$ intra-cluster sites by $\X$.  
So for each site in the original lattice $\x=\X+\tx$. In the reciprocal 
space, the points $\tx$ and $\X$ form lattices labeled by $\tk$ and 
$\K$ respectively with $K_\alpha=n_\alpha \cdot 2\pi/L$ and integer 
$n_\alpha$. Then $\k=\K+\tk$ (see Fig. \ref{divide_x_k}).

In the DCA, we first make the separation
\begin{equation}
\De=\frac{N}{N_c}\de_{\tk_1+\tk_2,\tk_3+\tk_4}\hspace{0.1cm}N_c
\de_{\K_1+\K_2,\K_3+\K_4}
\end{equation}
and then set $N/N_c\de_{\tk_1+\tk_2,\tk_3+\tk_4}\cong1$, so that
\begin{equation}
\label{eq:LDCA}
\De_{DCA}=N_c\de_{\K_1+\K_2,\K_3+\K_4}\,,
\end{equation}
which indicates that the momentum is partially conserved for transfers 
between the cells.\\ 

In this paper, we consider the approximation made through the 
substitution $\De \to \De_{DCA}$ in the compact and non-compact 
contributions to the thermodynamic potential.  Whenever the substitution 
is made, all internal legs are replaced by the coarse grained Green 
function defined by
\begin{equation}
\label{eq:cgGDCA}
\bG(\K,z)=\frac{N_c}{N}\sum_{\tk}G(\K+\tk,z)\, .
\end{equation} 
The corresponding estimate of the self energy will then necessarily be 
a function of $\K$ and (complex) frequency $z$. 
The DCA estimate of the lattice Green function is then given by
\begin{equation}
G(\k,z) =\frac{1}{z-\ep_\k+\mu-\Sigma_{DCA}(\K,z) } \,.
\label{G_DCA}
\end{equation}
It is of importance to note that by using Eq.~\ref{G_DCA} we have already 
made the approximation  
$\Sigma(k,z)=\Sigma_{DCA}(\K,z) + {\cal O}\big((\De k a)^2\big)$ 
where $\De k$ is the size of the coarse graining cell shown in Fig.
\ref{divide_x_k} and $a$ is the lattice constant (chosen as unity). We also
drop the frequency label from this point on for simplicity.

\begin{figure}
\epsfxsize=2.0in
\hspace*{2.0cm}\epsffile{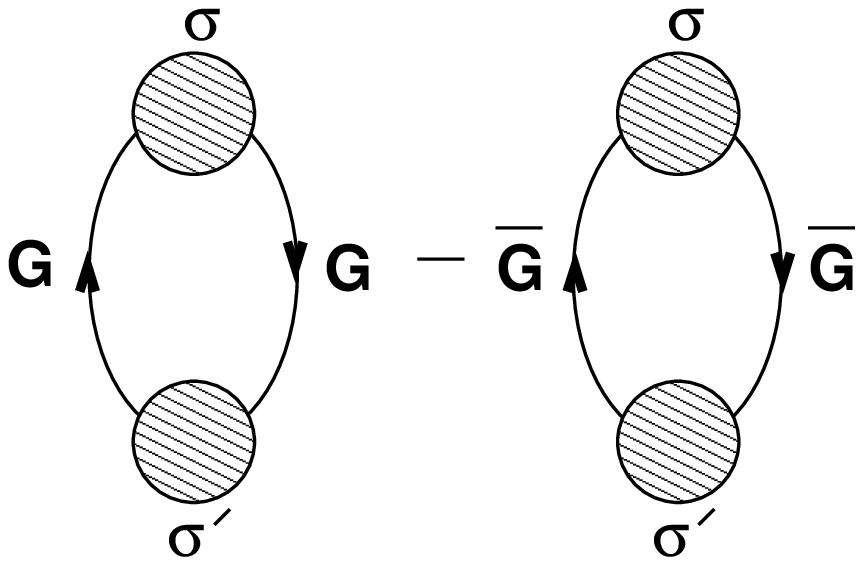}
\caption[a]{First correction by non-compact diagrams, 
$\delta^{(1)}[\De\Omega_{ncp}]$.
}
\label{noncomp-diff}
\end{figure}

\begin{figure}
\epsfxsize=2.7in
\hspace*{1.2cm}\epsffile{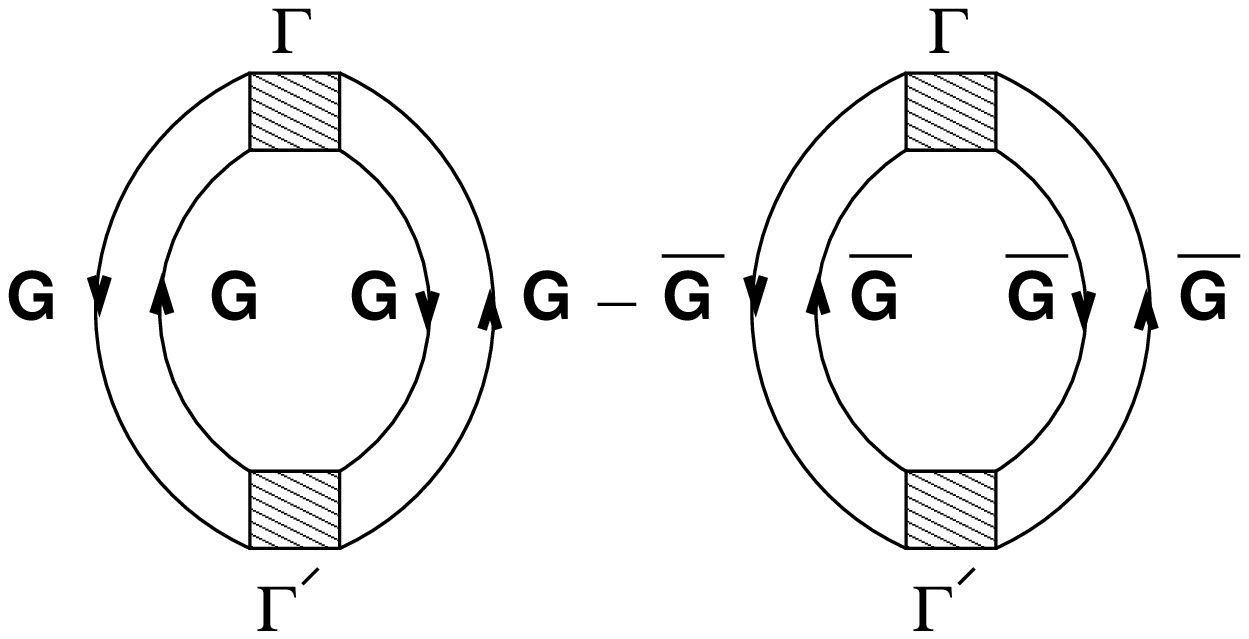} 
\caption[a]{First correction by compact diagrams, 
$\delta^{(1)}[\De\Omega_{cp}]$.
}
\label{comp-diff}
\end{figure}

\paragraph*{Coarse graining compact vs. non-compact diagrams}
We investigate the additional approximations associated with coarse 
graining in compact and non-compact diagrams.   To do this we will
consider diagrams with all legs coarse-grained, except those 
explicitly displayed in Figs.~\ref{noncomp-diff} and \ref{comp-diff}.  
Consider the first non trivial correction to the coarse grained non 
compact diagrams as illustrated in Fig.\ref{noncomp-diff}.  
\begin{eqnarray}
\delta^{(1)}[\De\Omega_{ncp}]\sim\frac{1}{N_c}\sum_{\K_1,\K_2}\sigma(\K_1,\K_2)
\sigma'(\K_1,\K_2)\int_{-\infty}^{\infty}\nonumber\\&&\hspace*{-7.5cm}
\int_{-\infty}^{\infty}d\epsilon_{1}d\epsilon_{2}G(\K_1,\epsilon_{1})
G(\K_2,\epsilon_{2})\big[\rho_{ncp}(\epsilon_{1},\K_1;\epsilon_{2},\K_2)- 
\nonumber\\&&\hspace*{-7.5cm}\bar{\rho}_{ncp}(\epsilon_{1},\K_1;\epsilon_{2},
\K_2)\big]\hspace*{0.1cm}\delta_{\K_2,\K_1}\,,
\label{delta_non_comp}
\end{eqnarray}   
where
\begin{eqnarray}
\rho_{ncp}(\epsilon_{1},\K_1;\epsilon_{2},\K_2)=
\frac{N_c}{N}\sum_{\tk_1,\tk_2}\delta(\epsilon_{_{1}}-\epsilon_{\K_1+\tk_1})
\times\nonumber\\&&\hspace*{-7.0cm}\delta(\epsilon_{_{2}}-\epsilon_
{\K_2+\tk_2})\hspace*{0.1cm}\delta_{\tk_2,\tk_1}\,,
\label{rho}
\end{eqnarray}  
and 
\begin{eqnarray}
\bar{\rho}_{ncp}(\epsilon_{1},\K_1;\epsilon_{2},\K_2)=
\frac{N_c^2}{N^2}\sum_{\tk_1,\tk_2}
\delta(\epsilon_{_{1}}-\epsilon_{\K_1+\tk_1})\times\nonumber\\&&
\hspace*{-7.0cm}\delta(\epsilon_{_{2}}-\epsilon_{\K_2+\tk_2})\,.
\label{rho_bar}
\end{eqnarray} 
In the above derivations we assumed that the self energy is $\tk$ 
independent and therefore the entire $\tk$ dependence of the
Green function is only through the  dispersion $\epsilon$ given by
\begin{equation}
\epsilon_{\K+\tk}=-2t/(2D)^{1/2}.\sum_{n=1}^{D}\cos(\K_{n}+\tk_{n})\,,
\label{disperssion}
\end{equation}
which is just the non interacting dispersion of the Hubbard model Hamiltonian 
with nearest neighbor hoppings and dimensionality $D$.\\
By the same token, in Fig.~\ref{comp-diff}, for the compact part with coarse 
grained $\Gamma$ and
$\Gamma'$ we have
\begin{eqnarray}
&& \delta^{(1)}[\De\Omega_{cp}]\sim\frac{1}{N_c^3}\sum_{\stackrel{\K_1,\K_2}
{_{\K_3,\K_4}}}
\delta_{\K_4,\K_1+\K_2-\K_3} \Gamma(\K_1,\K_2,\K_3,\K_4) \times\nonumber\\
&& 
\Gamma'(\K_1,\K_2,\K_3,\K_4)\int
_{-\infty}^{\infty}\int_{-\infty}^{\infty}\int
_{-\infty}^{\infty}\int_{-\infty}^{\infty}
d\epsilon_{1}d\epsilon_{2}d\epsilon_{3}d\epsilon_{4} \\
&& 
G(\K_1,\epsilon_{1})
G(\K_2,\epsilon_{2})G(\K_3,\epsilon_{3})G(\K_4,\epsilon_{4})
\,\big[\,\rho_{cp}(\epsilon_{1},\K_1;\epsilon_{2} \nonumber\\
&& 
,\K_2;\epsilon_{3},\K_3;\epsilon_{4},\K_4)
- \bar{\rho}_{cp}(\epsilon_{1},\K_1;\epsilon_{2},\K_2;\epsilon_{3},\K_3;
\epsilon_{4},\K_4)\hspace*{0.1cm}\big] \,, \nonumber
\label{delta_comp}
\end{eqnarray}
where 
\begin{eqnarray}
\rho_{cp}(\epsilon_{1},\K_1;\epsilon_{2},\K_2;\epsilon_{3},\K_3;\epsilon_{4},
\K_4)=\frac{N_c^3}{N^3}\sum_{\stackrel{\tk_1,\tk_2}{_{\tk_3,\tk_4}}}
\nonumber\\&&\hspace*{-6.8cm}\delta(\epsilon_{_{1}}-\epsilon_{\K_1+\tk_1})
\delta(\epsilon_{_{2}}-\epsilon_{\K_2+\tk_2})\delta(\epsilon_{_{3}}-\epsilon
_{\K_3 +\tk_3})\times\nonumber\\&&\hspace*{-6.8cm}\delta(\epsilon_{_{4}}-
\epsilon_{\K_4+\tk_4})\hspace*{0.1cm}\delta_{\tk_4,\tk_1+\tk_2-\tk_3}\,,
\label{rho_comp_ncg}
\end{eqnarray}
and
\begin{eqnarray}
\bar{\rho}_{cp}(\epsilon_{1},\K_1;\epsilon_{2},\K_2;\epsilon_{3},\K_3;
\epsilon_{4},\K_4)=\nonumber\\&&\hspace*{-5.0cm}\frac{N_c^4}{N^4}\sum_
{\stackrel{\tk_1,\tk_2}{_{\tk_3,\tk_4}}}\delta(\epsilon_{_{1}}-\epsilon_
{\K_1+\tk_1})\delta(\epsilon_{_{2}}-\epsilon_{\K_2+\tk_2})\times\nonumber
\\&&\hspace*{-5.0cm}\delta(\epsilon_{_{3}}-\epsilon_{\K_3+\tk_3})\delta
(\epsilon_{_{4}}-\epsilon_{\K_4+\tk_4})\,.
\label{rho_comp_cg}
\end{eqnarray}

\begin{figure}
\epsfxsize=3.3in
\hspace*{0.1cm}\epsffile{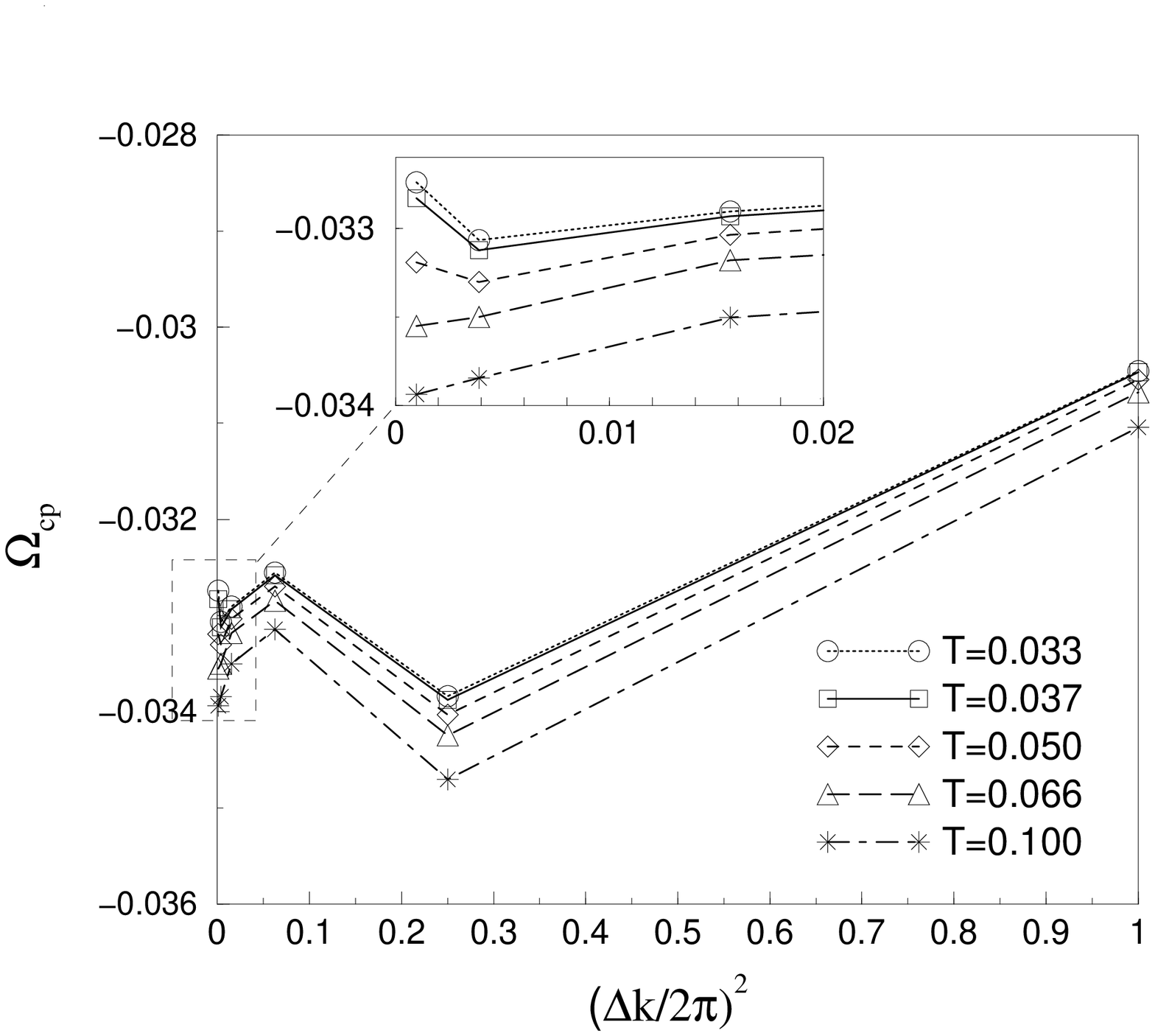} 
\caption[a]{The compact contribution of the thermodynamic potential
versus $\De k^2$ using FLEX method for $U=1.57$.}
\label{Phicp-vs-deltak^2}
\end{figure}
We now define the Fourier transforms of $\rho$ and $\bar{\rho}$
with respect to the energy arguments, e.g.
\begin{equation}
\Psi(s,\K)=\int_{-\infty}^{\infty}\rho(\epsilon,\K)e^{is\epsilon}d\epsilon \,.
\label{Psi(s)}
\end{equation}  
For the non-compact corrections, Eq.~\ref{Psi(s)} defines 
$\Psi_{ncp}(s_{1},\K_1;s_{2},\K_2)$ and 
$\bar{\Psi}_{ncp}(s_{1},\K_1;s_{2},\K_2)$, respectively.
In order to calculate the difference between $\Psi$ and $\bar{\Psi}$
in finite dimensions, we expand the dispersions and exponentials as 
functions of $\tk$ and keep the terms up to $\De k^2$. The difference 
$\delta\Psi_{ncp}$ for the non-compact diagrams becomes
\begin{eqnarray}
\delta\Psi_{ncp}(s_{1},\K_1;s_{2},\K_2)\approx
\frac{t^2}{3}\De k^2\times\nonumber\\&&\hspace*{-5.2cm}
\big[
\eta(\K_1+\K_2) -\eta(\K_1-\K_2) 
\big]\hspace*{0.05cm}s_1s_2\times\nonumber\\&&\hspace*{-5.2cm}
\exp\big[{i (s_1\epsilon_{\K_1}+s_2\epsilon_{\K_2})}\big]\hspace{0.05cm}\,,
\label{Psi-diff}
\end{eqnarray}
where $\eta(\K)=1/D\sum_{n}\cos(\K_n) \,.$
Reversing the Fourier transform then yields for the difference of $\rho_{ncp}$ 
and $\bar{\rho}_{ncp}$
\begin{eqnarray}
\delta\rho_{ncp}(\epsilon_{1},\K_1;\epsilon_{2},\K_2)\approx
-\frac{t^2}{3}\De k^2\times\nonumber\\&&\hspace*{-5.2cm}
\big[
\eta(\K_1+\K_2) -\eta(\K_1-\K_2) 
\big]\times\nonumber\\&&\hspace*{-5.2cm}
\frac{\partial}{\partial\epsilon_1}\delta(\epsilon_{1}-\epsilon_{\K_1})
\frac{\partial}{\partial\epsilon_2}\delta(\epsilon_{2}-\epsilon_{\K_2})\,.
\label{rho_noncomp_diff}
\end{eqnarray} 
Eq.~\ref{rho_noncomp_diff} demonstrates that the difference between coarse 
graining and not coarse graining in non-compact diagrams yields an error of 
order $\De k^2$. After tedious but straightforward calculation for the 
corresponding difference between coarse graining and not coarse graining in 
compact diagrams we obtain $\delta\rho_{cp} \sim {\cal O}(\De k^6)$.
Since $\De k=2\pi/L$, for large clusters this error becomes much smaller 
than the error in Eq.~\ref{rho_noncomp_diff}. 

\begin{figure}
\epsfxsize=3.2in
\hspace{0.3cm}\epsffile{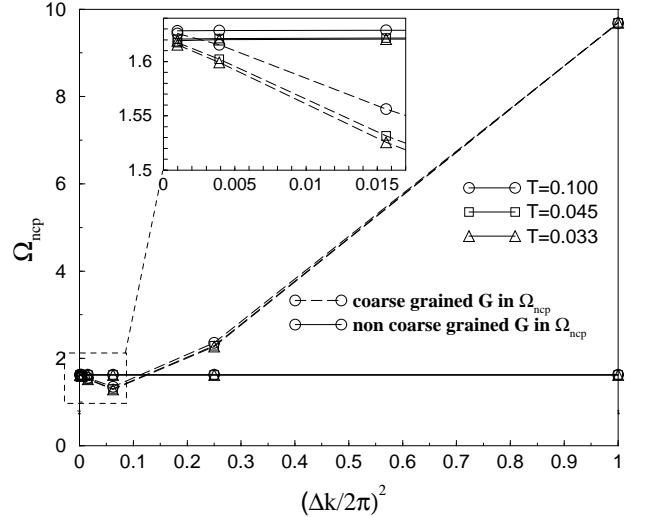} 
\vspace{0.5cm} 
\caption[a]{non-compact contributions of the
thermodynamic potential with and without coarse graining for $U=1.57$.}
\label{Phincp-vs-deltak^2}
\end{figure}
To illustrate the above point we simulate a two--dimensional Hubbard model 
(with local interaction $U$ and a nearest-neighbor hopping $t=1$)
using the Fluctuation Exchange Approximation (FLEX)\cite{bickers}.
We employ an elaborate subtraction scheme to correctly deal with
the high frequency behavior of the Green functions and FLEX 
potentials\cite{DHS}.
In Fig.~\ref{Phicp-vs-deltak^2} and Fig.~\ref{Phincp-vs-deltak^2} the 
compact and non-compact parts of $\De \Omega$ have been plotted
for the interaction $U=1.57$ for various cluster sizes $L$ and 
temperatures $T$.
In Fig. \ref{Phicp-vs-deltak^2}, it is readily seen that for the compact 
contribution with coarse grained Green functions the variation of
$\Omega_{cp}$ over the entire $\De k$ range is roughly 10\%.
In contrast,  as shown in Fig.~\ref{Phincp-vs-deltak^2}, the
difference between the non-compact contributions with and without coarse 
graining generates a deviations  of over 100\%. 
Note that at very low temperatures (inset in Fig.~\ref{Phicp-vs-deltak^2} ) 
the deviation from linearity  for T $<$ 0.066 is due to the correlation 
length exceeding the size of cluster $L$. Then, the DCA assumption of  
replacing $\Sigma(\k)$ by $\Sigma_{DCA}(\K)$ is no longer valid.

\paragraph*{Limit of Infinite Dimensions}
To make contact with the original derivations of the 
DMFA\cite{metzner-vollhardt,muller-hartmann}, we explore the 
differences between the compact and non-compact graphs in the
limit of many spatial dimensions, $D$.  For $D\rightarrow \infty$, we 
calculate $\De \rho$ for both compact and non-compact diagrams by 
simply expanding $\Psi$ and $\bar{\Psi}$ but this time with respect 
to 1/$D$ instead of $\De k^2$.  Letting $D\rightarrow \infty$ we can 
show that
   
\begin{eqnarray}
\lim_{D\rightarrow \infty} \Psi_{ncp}(s_{1},\K_1;s_{2},\K_2)
\approx\bar{\Psi}_{D\rightarrow \infty}\times\nonumber\\&&\hspace*{-6.0cm}
\rm exp\big(\frac{-t^2}{12}\De k^2 [\eta(\K_2-\K_1)-\eta(\K_1+\K_2)]s_1s_2\big
)\,,
\label{Psi-diff-inf}
\end{eqnarray}

\vspace{0.2cm}
Expanding the exponential in Eq. \ref{Psi-diff-inf}
\begin{eqnarray}
\lim_{D\rightarrow \infty} \Psi_{ncp}(s_{1},\K_1;s_{2},\K_2)\approx
\bar{\Psi}_{D\rightarrow \infty}\times\nonumber
\\&&\hspace*{-6.0cm}(1-\frac{t^2}{12}\De k^2
[\eta(\K_{2}-\K_{1})-\eta(\K_{1}+\K_{2})]\nonumber
\\&&\hspace*{-6.0cm}s_1s_2 +\ldots)\,,
\label{Psi-lin-inf}
\end{eqnarray}
and consequently
\begin{eqnarray}
\lim_{D\rightarrow \infty} 
\delta\rho_{ncp}(\epsilon_{1},\K_1;\epsilon_{2},\K_2)
\approx\frac{t^2}{12}\De k^2\times
\nonumber\\&&\hspace*{-6.0cm}
[\eta(\K_{2}-\K_{1})-\eta(\K_{1}+\K_{2})]\frac{\partial}
{\partial\epsilon_1}\frac{\partial}{\partial\epsilon_2}\nonumber
\\&&\hspace*{-6.0cm}\bar{\rho}_{D\rightarrow \infty}(\epsilon_{1},\K_1;
\epsilon_{2},\K_2)+{\cal O}( \De k^4)\,.
\label{rho-lin-inf}
\end{eqnarray}

\vspace{0.3cm}
However, for the difference in the compact diagrams as $D\rightarrow \infty$  
we find $\lim_{D\rightarrow \infty} \delta\Psi_{cp}=0$.
Thus, for all $\De k$, there are non-local corrections 
to non-compact diagrams while there are none to compact
diagrams as  $D\rightarrow \infty$. This result is consistent with what
M\"uller-Hartmann has shown for the DMFA \cite{muller-hartmann}. 
Regardless of whether the expansion parameter is 
$\Delta k$ or $1/D$, there is a fundamental difference between compact 
and non-compact diagrams which requires different treatments of each.   
Consequently, only the compact contributions should be formed from 
coarse-grained Green functions,  whereas non-compact graphs should be 
inferred from the appropriate Dyson equation.

Finally, even though we directly invoked the nearest neighbor hopping
dispersion in our algorithm, our arguments can be generalized to other 
models assuming that the Green function falls off exponentially as 
$G(r)\sim e^{-r/r_s}$ where $r_s$ is a characteristic length depending 
on dimensionality. It turns out that the compact diagrams fall off much 
faster due to having a larger number of Green functions relative to the 
non-compact ones which again explains why we coarse grain the Green 
functions in only the compact part of the free energy difference.
The arguments made here may also be applied to the DMFA
simply by taking $N_c=1$.  They also may be 
extended to other cluster approaches, such as the Molecular Coherent
Potential Approximation\cite{MCPA} and its formal equivalent for 
dynamical systems, the Molecular Cluster Dynamical Mean Field\cite{kotliar}.

\paragraph*{Conclusions} We justify one of the central underlying 
approximations of the dynamical cluster approximation.   We both 
analytically and numerically demonstrate that coarse graining 
the Green functions in non-compact diagrams results in a larger amount 
of error compared to that incurred in compact diagrams. Consequently, 
non-compact diagrams and their contribution to the thermodynamic 
potential are not coarse-grained.  The distinction between
non--compact and compact diagrams persists even in the 
limit of infinite dimensions. This concurs with previous work on
dynamical mean field theory and has implications for other cluster
approaches.\\

We would like to acknowledge Th.~Maier, R.~Jamei and A.~Voigt for very 
useful discussions. This work was supported by NSF grants DMR-0073308.

\end{document}